\documentclass[twocolumn, appendixfloats]{emulateapj}
\newcommand{\um}{\mbox{$\,\mu{\rm m}$}}

\newcommand{\etal}{et al.~}
\newcommand{\kms} {\mbox{\,km~s$^{-1}$}}

\newcommand{\CI}{[C\,{\sc i}]}

\newcommand{\LFIR}{\mbox{$L_{\rm FIR}$}}

\newcommand{\NII}{\mbox{[N\,{\sc ii}]}}
\newcommand{\CII}{\mbox{[C\,{\sc ii}]}}

\begin{document}

\shorttitle{CO\,(7$-6$), \CI\ 370\um\ and \NII\ 205\um\ Emission of BRI\,1335-0417}
\shortauthors{Lu et al.}

\title{CO (7$-$6), \CI\ 370\um\ and \NII\,205\um\ Line Emission 
       of the QSO BRI\,1335-0417 at Redshift 4.407}

\author{
Nanyao Lu\altaffilmark{1,2}, 
Tianwen Cao\altaffilmark{1,2,3},
Tanio D\'iaz-Santos\altaffilmark{4},
Yinghe Zhao\altaffilmark{5,6,7},
George C. Privon\altaffilmark{8,9,10},
Cheng Cheng\altaffilmark{1,2,11}, 
Yu Gao\altaffilmark{12},
C. Kevin Xu\altaffilmark{1,2},  
Vassilis Charmandaris\altaffilmark{13,14},
Dimitra Rigopoulou\altaffilmark{15}, 
Paul P. van der Werf\altaffilmark{16},
Jiasheng Huang\altaffilmark{1,2}, 
Zhong Wang\altaffilmark{1,2}, 
Aaron S. Evans\altaffilmark{17,18},
David B. Sanders\altaffilmark{19}
}

\altaffiltext{1}{National Astronomical Observatories, Chinese Academy of Sciences (CAS), Beijing 100012, China; nanyao.lu@gmail.com}
\altaffiltext{2}{China-Chile Joint Center for Astronomy, Camino El Observatorio 1515, Las Condes, Santiago, Chile}
\altaffiltext{3}{School of Astronomy and Space Science, University of Chinese Academy of Sciences, Beijing, China}
\altaffiltext{4}{Nucleo de Astronomia de la Facultad de Ingenieria, Universidad Diego Portales, Av. Ejercito Libertador 441, Santiago, Chile}
\altaffiltext{5}{Yunnan Observatories, Chinese Academy of Sciences, Kun- ming 650011, China}
\altaffiltext{6}{Key Laboratory for the Structure and Evolution of Celestial Objects, Chinese Academy of Sciences, Kunming 650011, China}
\altaffiltext{7}{Center for Astronomical Mega-Science, CAS, 20A Datun Road, Chaoyang District, Beijing 100012, China}
\altaffiltext{8}{Department of Astronomy, University of Florida, 211 Bryant Space Sciences Center, Gainesville, 32611 FL, USA}
\altaffiltext{9}{Departamento de Astronom\'ia, Universidad de Concepci\'on, Casilla 160-C, Concepci\'on, Chile}
\altaffiltext{10}{Pontificia Universidad Cat\'olica de Chile, Instituto de Astrofisica, Casilla 306, Santiago 22, Chile}
\altaffiltext{11}{Instituto de F\'isica y Astronom\'ia, Universidad de Valpara\'iso, Avda. 
 	         Gran Bretan\~a 1111, Valpara\'iso, Chile}
\altaffiltext{12}{Purple Mountain Observatory, CAS, Nanjing 210008, China}
\altaffiltext{13}{Department of Physics, University of Crete, GR-71003 Heraklion, Greece}
\altaffiltext{14}{IAASARS, National Observatory of Athens, GR-15236, Penteli, Greece}
\altaffiltext{15}{Department of Physics, University of Oxford, Keble Road, Oxford OX1 3RH, UK}
\altaffiltext{16}{Leiden Observatory, Leiden University, PO Box 9513, 2300 RA Leiden, The Netherlands}
\altaffiltext{17}{Department of Astronomy, University of Virginia, 530 McCormick Road, Charlottesville, VA 22904, USA}
\altaffiltext{18}{National Radio Astronomy Observatory, 520 Edgemont Road, Charlottesville, VA 22903, USA}
\altaffiltext{19}{University of Hawaii, Institute for Astronomy, 2680 Woodlawn Drive, Honolulu, HI 96822, USA}

\begin{abstract}
\noindent
We present the results from our Atacama Large Millimeter/submillimeter 
Array (ALMA) imaging observations of the CO\,(7$-$6), \CI\ 370\um\ 
(hereafter \CI) and \NII\ 205\um\ (hereafter \NII) lines and their 
underlying continuum emission of BRI\,1335-0417, an infrared bright
quasar at $z =$ 4.407. At the achieved resolutions of $\sim$1.1\arcsec\ 
to 1.2\arcsec\ (or 7.5 to 8.2 kpc), the continuum at 205 and 372\um\ 
(rest-frame), the CO\,(7$-$6), and the \CI\ emissions are at best 
barely resolved whereas the \NII\ emission is well resolved with 
a beam de-convolved major axis of 1.3\arcsec\,($\pm$0.3\arcsec) or
9\,($\pm$2) kpc.  As a warm dense
gas tracer, the CO\,(7$-$6) emission shows a more compact spatial
distribution and a significantly higher peak velocity dispersion 
than the other two lines that probe lower density gas, a 
picture favoring a merger-triggered star formation (SF) scenario 
over an orderly rotating SF disk.  The CO\,(7$-$6) data also indicate 
a possible QSO-driven gas outflow that reaches a maximum line-of-sight
velocity of 500 to 600\kms.  The far-infrared (FIR) dust temperature 
($T_{\rm dust}$) of 41.5\,K from a graybody fit to the continuum 
agrees well with the average $T_{\rm dust}$ inferred 
from various line luminosity ratios.  The resulting 
$L_{\rm CO(7-6)}/$\LFIR\ luminosity ratio is consistent with 
that of local luminous infrared galaxies powered 
predominantly by SF.  The $L_{\rm CO(7-6)}$-inferred SF rate is 
$5.1\,(\pm 1.5) \times 10^3\,M_{\odot}$\,yr$^{-1}$.  The system 
has an effective star-forming region of 1.7$^{+1.7}_{-0.8}$\,kpc 
in diameter and a molecular gas reservoir of $\sim$$5 \times 10^{11}\,M_{\odot}$.

\end{abstract}
\keywords{galaxies: active --- galaxies: ISM --- galaxies: star formation 
          --- infrared: galaxies --- ISM: molecules --- submillimeter: galaxies}

\vspace{-0.5in}
\section{INTRODUCTION} \label{sec1}

As more and more galaxies have been identified at high 
redshifts from recent deep photometric surveys, with 
some quasars (QSO) and other emission-line galaxies 
discovered at extremely high redshifts of $z > 7$ 
(e.g., Finkelstein et al. 2013; Watson et al. 2015; 
Venemans et al. 2017; Hu et al. 2017; Ba\~nados et 
al.~2018), how to effectively
characterize their star formation (SF) rate (SFR), SFR 
surface density ($\Sigma_{\rm SFR}$) and interstellar 
gas properties becomes an acute 
and yet challenging task. In particular, a direct 
measurement of $\Sigma_{\rm SFR}$ is difficult 
due to a requirement for high spatial resolution.

Among the high-$z$ galaxy samples studied so far (see 
Carilli \& Walter 2013 for a review), there is a population 
of the so-called sub-millimeter (sub-mm) galaxies (SMGs) at 
$z \gtrsim 2$, first identified in sub-mm bands
(Blain et al. 2002).
SMGs are among the brightest star-forming galaxies in 
the early Universe (Casey et al. 2014).
However, their large distances 
and dusty nature make it difficult to sufficiently 
reveal their internal SF structures at kpc to sub-kpc scale.  
As a result, it is still debatable whether 
their enormous bolometric luminosity is driven by a galaxy 
major merger, i.e., a scaled-up version of local 
ultra luminous infrared galaxies (ULIRGs; with a total 
8-1000\um\ luminosity, $L_{\rm IR} > 10^{12}\,L_{\odot}$;
e.g., Tacconi et al. 2006, 2008), or by a rotating, 
star-forming galaxy disk which is constantly fueled by 
the larger quantities of gas available at high-$z$ via 
an increased cold gas accretion rate (e.g., Agertz et 
al. 2009; Dekel et al. 2009; Dav\'e et al. 2010).

Lu et al. (2015) presented a simple spectroscopic approach
for simultaneously inferring the SFR, $\Sigma_{\rm SFR}$ 
and some molecular gas properties of a distant galaxy by 
measuring only the fluxes of the CO\,(7$-$6) line 
(rest-frame 806.652 GHz
or 372\um) and either the \NII\ line at 205\um\ (1461.134 GHz;
hereafter as \NII) or the \CII\ line at 158\um\ (1900.56 GHz;
hereafter as \CII).   For local luminous infrared galaxies 
(LIRGs; $L_{\rm IR} > 10^{11}\,L_{\odot}$) and ULIRGs, 
the CO\,(7$-$6) line luminosity, $L_{\rm CO(7-6)}$, can 
be used to infer the SFR of a galaxy with a $\sim$30\% 
accuracy, irrespective of whether the galaxy hosts an active
galactic nucleus (AGN; Lu et al. 2014, 2015, 2017a; 
Zhao et al. 2016a).  Furthermore, the steep anti-correlation
between the \NII/CO\,(7$-$6) (or \CII/CO\,(7$-$6)) luminosity
ratio and the rest-frame far-infrared (FIR) color, 
$C(60/100)$ ($\equiv$ $f_{\nu}(60\um)/f_{\nu}(100\um)$), 
can be used to estimate $C(60/100)$ or the dust temperature 
$T_{\rm dust}$ (Lu et al. 2015).   $C(60/100)$ is in turn 
related to $\Sigma_{\rm SFR}$  (Liu et al. 2015; Lutz et al. 2016). 
Alternatively, one can estimate $C(60/100)$ from  
the \CI/CO\,(7$-$6) luminosity ratio (Lu et al. 2017a; also 
see Appendix A), where \CI\ refers to the fine-structure 
transition ($^3P_2 \rightarrow$$^3P_1$) at 370\um\ (809.342 GHz)
of neutral carbon.  The frequency separation between 
the \CI\ and CO\,(7$-$6) lines is only $2.69/(1+z)$\,GHz 
at high redshift $z$. This 
greatly increases observational efficiency as both lines 
can fit within the same sideband of the Atacama Large 
Millimeter/submillimeter Array (ALMA; Wootten \& Thompson 2009).  
These indirect approaches to estimating $\Sigma_{\rm SFR}$ are 
useful at high redshifts, where it is often challenging to 
resolve a galaxy in the FIR/sub-mm.

In addition, these gas cooling lines probe different interstellar
gas phases.  With a critical density ($n_{\rm crit}$) of 
$\sim$$10^{5}$\,cm$^{-3}$ and an excitation temperature 
($T_{\rm ex}$) of $\sim$150\,K (Carilli \& Walter 2013),  
the CO\,(7$-$6) line probes dense and warm molecular gas 
that is energetically associated with and in proximity of 
current or very recent SF activities (Lu et al. 2014, 2017a). 
The \CI\ line has $n_{\rm crit} \sim 10^3$\,cm$^{-3}$, similar
to that of the CO\,(1$-$0) line (Carilli \& Walter 2013). 
Recent observations suggest that
this line traces general molecular gas as the CO\,(1$-$0) line 
does (e.g., Papadopoulos et al. 2004a, 2004b; Lu et al. 2017a;  
Jiao et al. 2017).   The \NII\ line has a low $n_{\rm crit}
\sim 50$\,cm$^{-3}$ (Carilli \& Walter 2013) for collisional 
excitation with electrons.  This line probes mainly diffuse, 
hot ionized gas (Zhao et al. 2016b; D\'iaz-Santos et al. 2017).

A hallmark of a galaxy major merger is the molecular gas funneled 
into the inner region of the merging galaxies (Solomon \& Sage 1988;
Sanders et al. 1988; Scoville et al. 1989; Sanders \& Mirabel 1996;
Solomon et al. 1997; Downes \& Solomon 1998; Gao \& Solomon 1999;
Evans et al. 2002) as a result of gravitational torque during 
the merger (Barnes \& Hernquist 1996; Hopkins et al. 2009).  
Furthermore, the non-axisymmetric tidal force also leads to gas 
turbulences and shocks that compress the gas into higher densities
(e.g., Bournaud et al. 2011).  Recent cosmological simulations 
of galaxy mergers at high $z$ (e.g., Sparre \& Springel 2016) 
suggest that different gas phases have different spatial 
distributions, with (a) star-forming dense gas in the inner region 
of the merging galaxies, and (b) diffuse hot ionized gas extending
to large radii.  The region between (a) and (b) is dominated 
by (c) gas of intermediate densities.  In this merger scenario, 
the CO\,(7$-$6) line traces predominantly the gas phase (a); 
the \NII\ would be particularly sensitive to the gas phase (b).
Likely, the \CI\ could have a significant contribution from 
the gas phase (c). Such spatial scale differences are indeed
observed in some local advanced mergers between, for example,
CO\,(6$-$5) and a low-$J$ CO line such as CO\,(1$-$0) or 
CO\,(2$-$1) (e.g., Xu et al. 2014, 2015).
On the other hand, in an orderly rotating disk SF scenario, 
the SF occurs in dense blobs embedded in the gaseous disk.
As a result, both CO\,(7$-$6) and \CI\ may reflect the same 
disk geometry and kinematics. 

In this paper we present the results from our ALMA observations 
of the CO\,(7$-6$), \CI\ and \NII\ line emission of BRI 1335-0417, 
an infrared luminous QSO at $z = 4.407$ (Storrie-Lombardi 
et al. 1994), as part of our continued effort to expand the number 
of $z > 4$ galaxies with CO\,($7-6$), \CI,  \NII, and \CII\ 
detections.  The host galaxy system of this unlensed QSO 
(Storrie-Lombardi et al. 1996) is dusty and gas rich (Omont et al. 1996;
Guilloteau et al. 1997; Carilli et al. 1999, 2002;  
Benford et al. 1999; Yun et al. 2000; Wagg et al. 2014; 
Jones et al. 2016) and is likely going through a major merger 
involving two galaxy progenitors roughly along the north-south 
direction,  separated by about 0.6\arcsec\ ($\sim$4 kpc; Riechers 
et al. 2008).  The QSO, likely hosted by the dominant southern 
galaxy member, has an 
estimated black hole mass of $\sim$$6\times 10^9\,M_{\odot}$ 
(Shields et al. 2006).  The detection 
of the CO\,(5$-$4) emission (Guilloteau et al. 1997) indicates 
a large amount of warm and dense molecular gas associated with 
the on-going intense SF.  The ALMA data presented here allow 
for not only some quantitative characterization of the SF and gas 
properties, but also further testing the galaxy merger 
scenario for this system, which is caught at the stage of active 
assembling of a massive galaxy and rapid growth of the central 
massive black hole when the Universe was only $\sim$1.4 billion 
years old.

In the remainder of this paper, we describe our observations, data 
reduction and results in \S2, analyze the observed line and dust 
continuum emission, quantify the SF and gas properties, and 
discuss merger-dominated SF scenario in \S3,  and finally 
summarize our results in \S4.  
Throughout the paper we use a flat cosmology with $\Omega_M = 0.27$, 
$\Omega_{\Lambda} = 0.73$ and $H_0 = 71$\,\kms\,Mpc$^{-1}$. At $z = 
4.407$, the luminosity distance is 40,993 Mpc and 1\arcsec\ corresponds 
to 6.8 kpc.


\begin{deluxetable*}{ll}
\tabletypesize{\footnotesize}
\tablenum{1}
\tablewidth{0pt}
\tablecaption{Observed and Derived Parameters$^a$}
\tablehead{
\colhead{\hspace{-0.50in}Parameter} & \colhead{\hspace{-0.6in}Value}}
Observations: \\
\ \ \ ALMA beam ($\nu_{\rm obs} \approx$ 149 GHz) \hspace{0.0in}    & \hspace{0.0in} (1.2\arcsec$\times$0.9\arcsec, 82\arcdeg) \\
\ \ \ ALMA beam ($\nu_{\rm obs} \approx$ 270 GHz)                   & \hspace{0.0in} (1.1\arcsec$\times$0.8\arcsec, 64\arcdeg)  \\
\\
Continuum: \\
\ \ \ Gaussian fit position$^b$ (J2000)   & \hspace{0.0in} (13$^{\rm h}$38$^{\rm m}$03{\fs}419, -4{\arcdeg}32{\arcmin}35{\farcs}06) \\
\ \ \ Gaussian fit size$^c$ (149 GHz)     & \hspace{0.0in} 1.20\arcsec\,($\pm$0.04\arcsec)$\times$0.96\arcsec\,($\pm$0.03\arcsec), 89\arcdeg($\pm$6\arcdeg) \\
\ \ \ $S_{\nu}$ (149 GHz) (mJy)           & \hspace{0.0in} 1.17 ($\pm 0.07$)          		    \\
\ \ \ Gaussian fit size$^c$ (270 GHz)     & \hspace{0.0in} 1.18\arcsec\,($\pm$0.01\arcsec)$\times$0.92\arcsec\,($\pm$0.01\arcsec), 61\arcdeg($\pm$1\arcdeg) \\
\ \ \ $S_{\nu}$ (270 GHz) (mJy)           & \hspace{0.0in} 9.03 ($\pm 0.11$)             		    \\
CO\,(7$-$6)$^d$: \\
\ \ \ Gaussian fit position (J2000)    & \hspace{0.0in} (13$^{\rm h}$38$^{\rm m}$03{\fs}420, -4{\arcdeg}32{\arcmin}35{\farcs}05) \\
\ \ \ Gaussian fit size$^c$            & \hspace{0.0in} 1.34\arcsec\,($\pm$0.06\arcsec)$\times$0.97\arcsec\,($\pm$0.04\arcsec), 67\arcdeg($\pm$5\arcdeg)   \\
\ \ \ Central frequency (GHz)          & \hspace{0.0in} 149.179 ($\pm 0.003$)                          \\
\ \ \ FWHM (\kms)                      & \hspace{0.0in} 341  ($\pm 11$)                              \\
\ \ \ Flux (Jy\kms)                    & \hspace{0.0in} 3.08 ($\pm 0.11$)                             \\
{\CI}$^d$: \\
\ \ \ Gaussian fit position (J2000)     & \hspace{0.0in} (13$^{\rm h}$38$^{\rm m}$03{\fs}424, -4{\arcdeg}32{\arcmin}35{\farcs}03) \\
\ \ \ Gaussian fit size$^c$             & \hspace{0.0in} 1.38\arcsec\,($\pm$0.15\arcsec)$\times$1.06\arcsec\,($\pm$0.12\arcsec), 68\arcdeg($\pm$17\arcdeg)   \\
\ \ \ Central frequency (GHz)           & \hspace{0.0in} 149.690 ($\pm 0.004$)                           \\
\ \ \ FWHM (\kms)                       & \hspace{0.0in} 314 ($\pm 16$)                               \\
\ \ \ Flux (Jy\kms)                     & \hspace{0.0in}  1.04 ($\pm 0.09$)                             \\
\NII$^d$ (total): \\
\ \ \ Gaussian fit position (J2000)                 & \hspace{0.0in} (13$^{\rm h}$38$^{\rm m}$03{\fs}412, -4{\arcdeg}32{\arcmin}35{\farcs}09) \\
\ \ \ Gaussian fit size$^c$           & \hspace{0.0in}  1.70\arcsec\,($\pm$0.32\arcsec)$\times$0.92\arcsec\,($\pm$0.14\arcsec), 49\arcdeg($\pm$9\arcdeg)   \\
\ \ \ Central frequency (GHz)          & \hspace{0.0in} 270.24  ($\pm 0.03$)                         \\
\ \ \ FWHM  (\kms)                     & \hspace{0.0in} 603  ($\pm 64$)		                    \\
\ \ \ Flux (Jy\kms)                    & \hspace{0.0in} 1.95 ($\pm 0.23$)                             \\
\NII$^d$ (3-Gaussian fit)$^e$:\\
\ \ \ Frequency (core) (GHz)          & \hspace{0.0in} 270.27 ($\pm 0.01)$			   \\
\ \ \ FWHM (core) (\kms)              & \hspace{0.0in} 238  ($\pm 31$)		                    \\
\ \ \ Flux (core) (Jy\kms)            & \hspace{0.0in} 1.04 ($\pm 0.14$)                           \\
\ \ \ Frequency (red side) (GHz)      & \hspace{0.0in} 269.93 ($\pm 0.05)$			   \\
\ \ \ FWHM (red side) (\kms)          & \hspace{0.0in} 394  ($\pm 109$)		           \\
\ \ \ Flux (red side) (Jy\kms)        & \hspace{0.0in} 0.62 ($\pm 0.18$)                           \\
\ \ \ Frequency (blue side) (GHz)     & \hspace{0.0in} 270.58	($\pm 0.02)$			    \\
\ \ \ FWHM (blue side) (\kms)         & \hspace{0.0in} 147  ($\pm 49$)		                    \\
\ \ \ Flux (blue side) (Jy\kms)       & \hspace{0.0in} 0.34 ($\pm 0.12$)                             \\
\\
Line and continuum luminosities: \\
\ \ \ $L_{\rm [NII], total}/L_{\odot}$           & \hspace{0.0in} $9.2\,(\pm 1.1) \times 10^8$    \\ 
\ \ \ $L_{\rm [NII], core}/L_{\odot}$            & \hspace{0.0in} $4.9\,(\pm 0.7) \times 10^8$    \\   
\ \ \ $L_{\rm CO(7-6)}/L_{\odot}$      		 & \hspace{0.0in} $8.0\,(\pm 0.3) \times 10^8$    \\ 
\ \ \ $L_{\rm [CI]}/L_{\odot}$      		 & \hspace{0.0in} $2.7\,(\pm 0.2) \times 10^8$    \\ 
\ \ \ $L_{\rm [CII]}^f/L_{\odot}$      		 & \hspace{0.0in} $1.6\,(\pm 0.3) \times 10^{10}$    \\ 
\ \ \ $L_{\rm FIR}^g/L_{\odot}$      		 & \hspace{0.0in} $2.0\,(\pm 0.1) \times 10^{13}$    \\ 
\enddata
\tablenotetext{a}{ALMA flux uncertainties cited do not include the absolute calibration uncertainty likely at 
		  $\sim$10\%.  The uncertainties of the continuum flux and source size based on a 2d Gaussian 
 		  fit to an image were estimated following the prescription in Condon (1997) for correlated noise 
              cases. The uncertainties for the parameters from a Gaussian fit to a spectrum were estimated 
		  following the formulae in Lenz \& Ayres (1992).}
\tablenotetext{b}{Taken from the 2d Gaussian fit to the 270 GHz continuum image which has a higher S/N between
		  the two continuum images.  The measured position 
		  difference between the two continuum bands is 0{\farcs}09. The typical astrometric accuracy of 
		  of our ALMA observations is $\sim$0.1\arcsec.}
\tablenotetext{c}{FWHM major and minor axes, followed by the major axis PA (N to E), from the 2d Gaussian 
		  fit.}
\tablenotetext{d}{Spectrum extracted from within an elliptical aperture that resembles the FWHM ellipse of
	 	  the 2d Gaussian fit to the frequency-integrated line image, but with the major and minor axes 
              each stretched by a factor of $n=2.5$.  For the CO\,(7$-$6) and \CI\ lines, we used a common 
              aperture referenced to the 2d Gaussian fit to the \CI\ image.} 
\tablenotetext{e}{These are 3 Gaussian components from the fit to the \NII\ spectrum in Fig. 3c.}
\tablenotetext{f}{Based on the \CII\ flux in Wagg et al. (2010).}
\tablenotetext{g}{The quoted luminosity error reflects how much the fitted SED-based $L_{\rm FIR}$ varies 
                  if $T_{\rm dust}$ changes from 41.5\,K by $\pm$2\,K.}
\end{deluxetable*}

\section{Observations and Results} \label{sec2}

\subsection{Observations and Data Reduction} \label{sec2.1}

The CO\,(7$-$6)/\CI\ observation of BRI 1334-0417 was carried 
out in the ALMA Band 4 in the time division mode (TDM) in two 
equal-duration runs on March 4 and 19, 2016, respectively.
The observation utilized 39 of the 12-meter antennas with baselines 
ranging from 15 to 460 meters.  The effective total on-target 
integration is 607 seconds. The separate ALMA band-6 \NII\ 
observation was conducted in TDM in one session on 
January 4, 2016, with an on-target integration of 303 seconds.
A total of 40 of the 12-meter antennas were used, with baselines 
from 15 to 331 meters.  In each observation, one of the 4 
spectral windows (SPWs), each of 1875 MHz wide, was used to 
cover the redshifted line(s),  and the other 3, tuned 
at some nearby frequencies on both sides of the spectral 
line, were used for the continuum measurement.  Each SPW 
has 128 channels with a channel width of 15.625 MHz.  
The effective spectral resolution is 31.25 MHz, equivalent 
to 63 and 35\kms\ for the CO\,(7$-$6)/\CI\ and \NII\ 
observations, respectively. The CO\,(7$-$6) and \CI\ lines
are covered by the same SPW as the two lines are separated
by only 0.50 GHz (= 2.69 GHz/(1+$z$)).  The pointing, phase, 
bandpass and flux calibrations were based on Ganymede, 
J1116+0829,  J1332-0509 and J1337-1257 in the 
CO\,(7$-$6)/\CI\ observation and on J1337-1257 and J1332-0509
in the \NII\ observation.

The data reduction was carried out with the Common Astronomy Software
Applications (CASA) 4.5.3 and the final images were cleaned using 
the Briggs weighting with the parameter ``robust'' = 0, resulting
in a synthesized beam of full width at half maximum (FWHM) of 
1.2\arcsec$\times$0.9\arcsec\ (1.1\arcsec$\times$0.8\arcsec) at 
a position angle (PA; N to E) of 82\arcdeg\ (64\arcdeg) for 
the CO\,(7$-$6)/\CI\ (\NII) observation.  The r.m.s.~noise in the final 
continuum image is $\sim$38 (57)\,$\mu$Jy\,beam$^{-1}$ from 
the CO\,(7$-$6) (\NII) observation.  For the spectral cube data, 
the continuum was removed using the CASA function ``uvcontsub''
of order $= 1$. We further reduced the channel width of the final 
spectral cube to 100\kms, resulting in an r.m.s.~noise of $\sim$0.29 
(0.37)\,mJy\,beam$^{-1}$ for 
the CO\,(7$-$6)/\CI\ (\NII) data.

\begin{figure*}
\vspace{-0.5in}
\centering
\includegraphics[width=1.0\textwidth, bb=0 0 1008 1008]{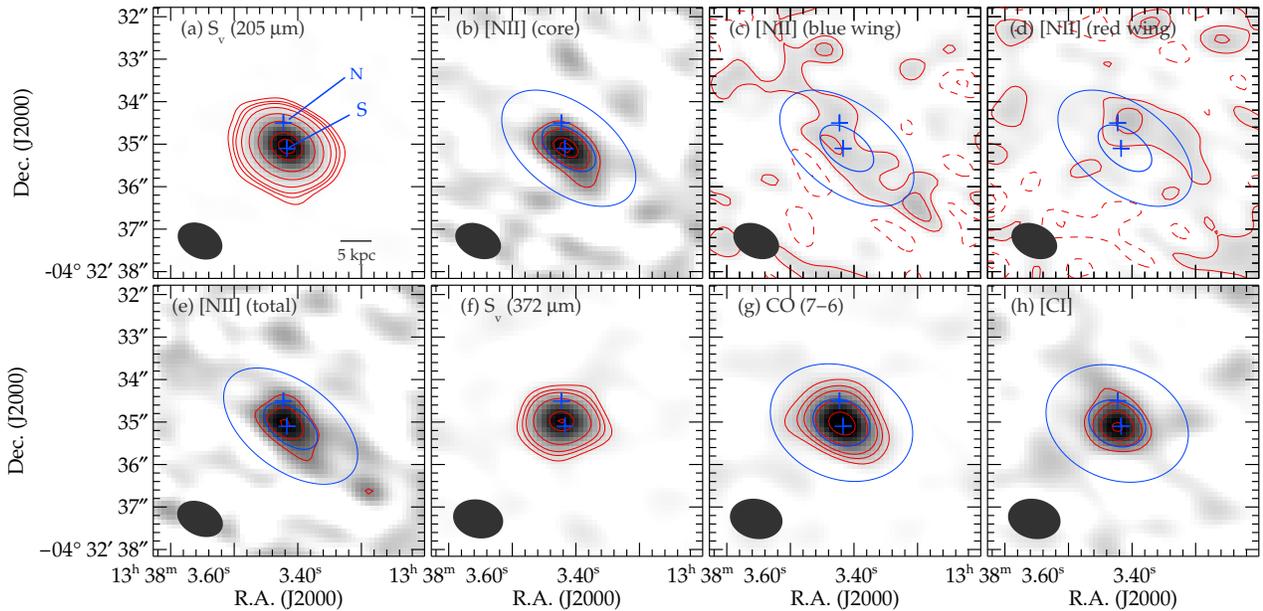}
\vspace{-2.8in}
\caption{
Surface brightness images: (a) the continuum at $\sim$205\um\ (rest frame); 
the \NII\ line emission integrated over (b) the core frequency range 
from 270.021 to 270.472 GHz (243 to -257\kms), 
(c) the blue wing (270.472 to 270.742 GHz or -257 to -557\kms), 
(d) the red wing (269.751 to 270.021 GHz or 543 to 243\kms), 
or (e) the full frequency range from 269.751 to 270.742 GHz;  
(f) the continuum at $\sim$372\um\ (rest frame); (g) the CO\,(7$-$6) 
emission integrated from 148.902 to 149.449 GHz (543 to -557\kms);
and (h) the \CI\ 370\um\ emission integrated from 149.499 
to 149.847 GHz (320 to -377\kms).  
In each panel, the image and the red contours refer 
to the same data; the image stretch is from 0 to the highest contour 
plotted except for panels (c) and (d), in which the gray scales are 
the same as in panel (b).  All the contours start at 3$\sigma$, except
for those in panels (c) and (d).   In units of mJy\,beam$^{-1}$, 
the contours are shown at [3, 5, 7, 12, 24, 48, 96, 120] $\times$ 0.057 in (a)
or at [3, 5, 7, 12, 24, 27.6] $\times$ 0.038 in (f).  In units of
Jy\,\kms\,beam$^{-1}$, the contours in (b), (e), (g) and (h)
are plotted at [3, 5, 7, 7.7] $\times$ 0.11,
[3, 5, 6.4] $\times$ 0.17, [3, 5, 7, 12, 19] $\times$ 0.12, 
and [3, 5, 7, 9] $\times$ 0.092, respectively. The contours in 
both (c) and (d) are at [-2, -1, 1, 2] $\times \sigma$, where 
$\sigma = 0.08$ Jy\,\kms\,beam$^{-1}$, and the contours of 
negative values are shown in dashed line.
The black ellipse at the lower left corner in each panel indicates
the relevant (FWHM) synthesized ALMA beam.  The blue ellipses in 
each line image indicate respectively the two apertures used for 
extracting the 1d spectra in Fig. 3.   The two blue crosses in 
each panel mark the locations of sources N and S (see the text).  
The 5 kpc scale is shown in panel (a).
}
\label{Fig1}
\end{figure*}

\subsection{Results} \label{sec2.2}

We show in Fig.~1 the two continuum images as well as the 
frequency-integrated line images.  For the \NII\ emission, 
the spatially integrated line profile shows possible 
high-velocity components (as analyzed in \S3.3). We therefore
include in Fig.~1 four separate frequency-integrated images 
for the \NII\ emission:  a ``core'' image (Fig. 1b) that 
encompasses the main emission component, two images 
representing the blue (Fig. 1c) and red (Fig. 1d) wing 
components, respectively, and a ``total'' image (Fig. 1e) 
that was integrated over the full frequency range. 
In each panel, the two blue crosses mark the respective 
positions of the two main gas components 
resolved in the Very Large Array (VLA) CO\,(2$-$1) map 
in Riechers et al. (2008). 
These two sources (hereafter referred to as sources N 
and S) are separated by $\sim$200\kms\ along the line-of-sight
and by $\sim$0.6\arcsec\ ($\sim$4 kpc) spatially.  
As shown by Riechers 
et al, these individual sources are rich in molecular gas and 
likely represent the progenitors in an on-going major galaxy merger.

\begin{figure*}
\vspace{-0.5in}
\centering
\includegraphics[width=1.0\textwidth, bb=0 0 958 1008]{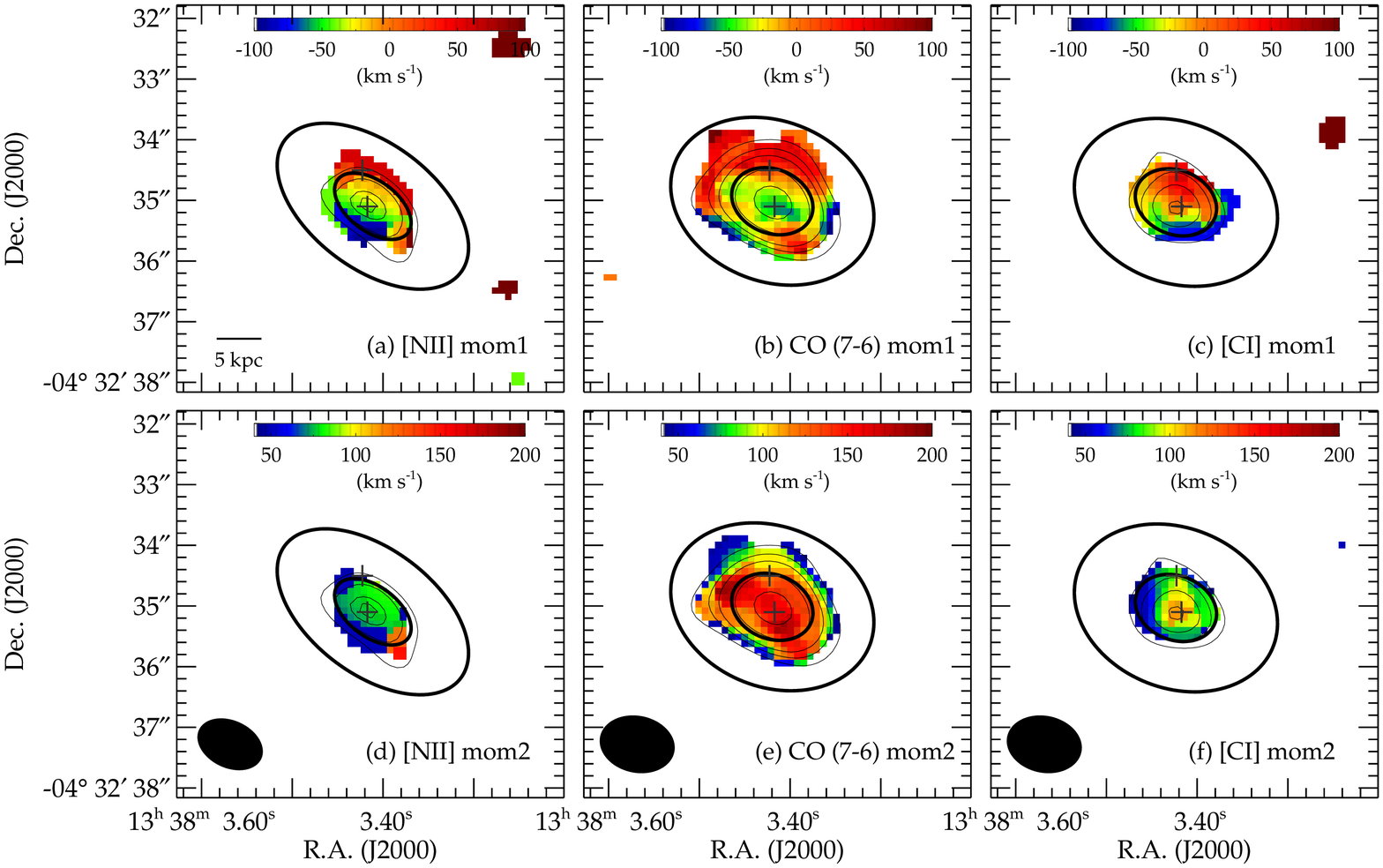}
\vspace{-2.65in}
\caption{First moment (top row) and second moment (bottom row) images of the \NII,  
CO\,(7$-$6) and \CI\ lines, respectively. Only channel pixels (spaxels) 
above 2.5$\sigma$ were used in deriving these images, where $\sigma$ is 
the r.m.s. spaxel-to-spaxel noise.  The contours overlaid are that from 
the corresponding panel in Fig. 1. For the \NII\ line, the contours form 
Fig. 1b are shown here.  For clarity, the same crosses and the elliptical
apertures as in Fig. 1 are also plotted here, but in thick black lines.
}
\label{Fig2}
\end{figure*}

We fit a 2d Gaussian function to each of the images in Fig. 1, except
for the image panels (b) through (d).  The resulting central position, 
FWHM major and minor axes, and PA are given in Table 1.  Based on these
results, the continuum emission at 205 and 372\um\ (rest-frame) are 
both unresolved or just barely resolved, as well as the CO\,(7$-$6) 
and \CI\ emission. In contrast,  the \NII\ emission appears to be 
well resolved, with an ALMA beam-deconvolved major axis ($d_{\rm deconv}$)
of 1.3\arcsec\,($\pm$0.4\arcsec) (9$\pm$3 kpc) as discussed in more 
detail in \S3.3.

Fig. 2 compares the moment-1 and moment-2 images of the emission lines 
using the CASA function ``immoments'' on all the channel data points above 
2.5$\sigma$, where $\sigma$ is the pixel-to-pixel noise per frequency 
channel (i.e., spaxel).  In addition, we used only the channels free from 
contamination by another spectral line.   Therefore, these surface 
brightness-weighted images show mainly the inner galaxy.   
All moment-1 images indicate a similar overall velocity field, with 
a roughly north-south velocity gradient of up to $\sim$150\kms. 
There are subtle differences between the lines: for example, the velocity 
field increases to $+$50\kms\ towards a south-west patch in the moment-1
image of the CO\,(7$-6$) line. However, this patch completely disappears if 
the spaxel S/N cutoff is raised to S/N $>3.5$. Therefore, these subtle 
differences are of low detection significance.  In contrast, the moment-2 
images reveal significant differences between the lines:  the \NII\ line 
shows a smoothly varying velocity dispersion up to 80\kms\ over 
the inner galaxy except for a higher value in a region at $\sim$1.1\arcsec\
south-west of source S.  Upon a closer examination of the spectral cube, 
this region of apparently higher velocity dispersion is likely caused 
by a redshifted (at -185\kms) signal detected only at S/N $\sim 3$, 
which has no clear corresponding signal in the other 
two lines.  By comparison, the CO\,(7$-$6) emission shows a more patchy 
velocity dispersion field with an elongated area at PA $\sim 45$\arcdeg\ 
showing higher velocity dispersions of 150 to 180\kms. This 
patchiness remains identifiable until the spaxel S/N cutoff is raised to S/N 
$>5$ in constructing the moment images.
For the \CI\ line, a peak velocity dispersion of $\sim$120\kms\ is 
seen around source S.  Therefore, the dense molecular gas traced 
by CO\,(7$-$6) is effectively subject to a higher turbulent velocity
field than the gas component traced by either \CI\ or \NII. It is thus 
unlikely that the CO\,(7$-$6) and \CI\ line emission arise from 
the same region.  We analyze this further in \S3.2 and discuss its 
physical implications in \S3.6.

Fig. 3 shows the extracted 1d spectra using an elliptical aperture
of the size index $n = 2.5$ (upper panels) or $n = 1.0$ (lower 
panels), where the index $n$ refers to the common stretching factor 
for the major and minor axes of the aperture relative to that of a 
reference aperture ($n = 1$), with the aperture PA always fixed.
Therefore, the two apertures used in Fig. 3 
differ in their major and minor axes by a common scaling factor 
of $2.5/1.0$.  For the spectra encompassing the CO\,(7$-$6) and \CI\ 
lines (left panels in Fig. 3), the aperture of $n = 1.0$ is defined
by having its FWHM major and minor axes and PA set to the corresponding
values in the 2d Gaussian fit result for the \CI\ line in Table 1; 
for the \NII\ spectrum (right panels in Fig. 3), it is set to the 2d 
Gaussian fit result for the total \NII\ image in Table 1.  These
$n=1.0$ and $n=2.5$ apertures are also plotted in Figs. 1 and 2.

We fit a Gaussian function to each spectral line in Fig. 3 and 
the results are shown by the black curves.  The \NII\ 
line profile in panel (c) apparently shows wing features 
on both sides, indicating possible gas emission at velocities 
of about $\pm$350\kms.  We therefore also fit 3 Gaussian 
components simultaneously to the spectrum, shown by the blue curves. 
The \NII\ wing features become weaker from 
the top-right panel to the bottom-right panel in Fig. 3, 
indicating that the high-velocity \NII\ emission is distributed
over a wide sky area.  Their spatial morphology,  shown in 
panels (c) and (d) in Fig. 1, indicates two narrow and long 
stretches that is apparently reminiscent of tidal tails. 
The presence of tidal gas tails would firmly establish 
the on-going galaxy major merger scenario (Toomre \& Toomre 1972). 
However, this requires further observational confirmation as 
these high-velocity \NII\ features are detected only at S/N
$\approx 2$$-$3.

The velocities of the \NII\ wing components are also indicated 
for the CO\,(7$-$6) and \CI\ lines in the left panels in Fig. 3. 
We detect some emission 
excess on both wings of the main Gaussian fit to the CO\,(7$-$6) 
emission. As our analysis in \S3.1 shows, this high-velocity 
CO\,(7$-$6) emission is likely associated with the dominant 
source S, which is probably the host of the QSO.
\vspace{0.1in}

\begin{figure*}
\begin{center}
\begin{tabular}{cc}
\includegraphics[width=0.40\textwidth, bb=200 250 410 718]{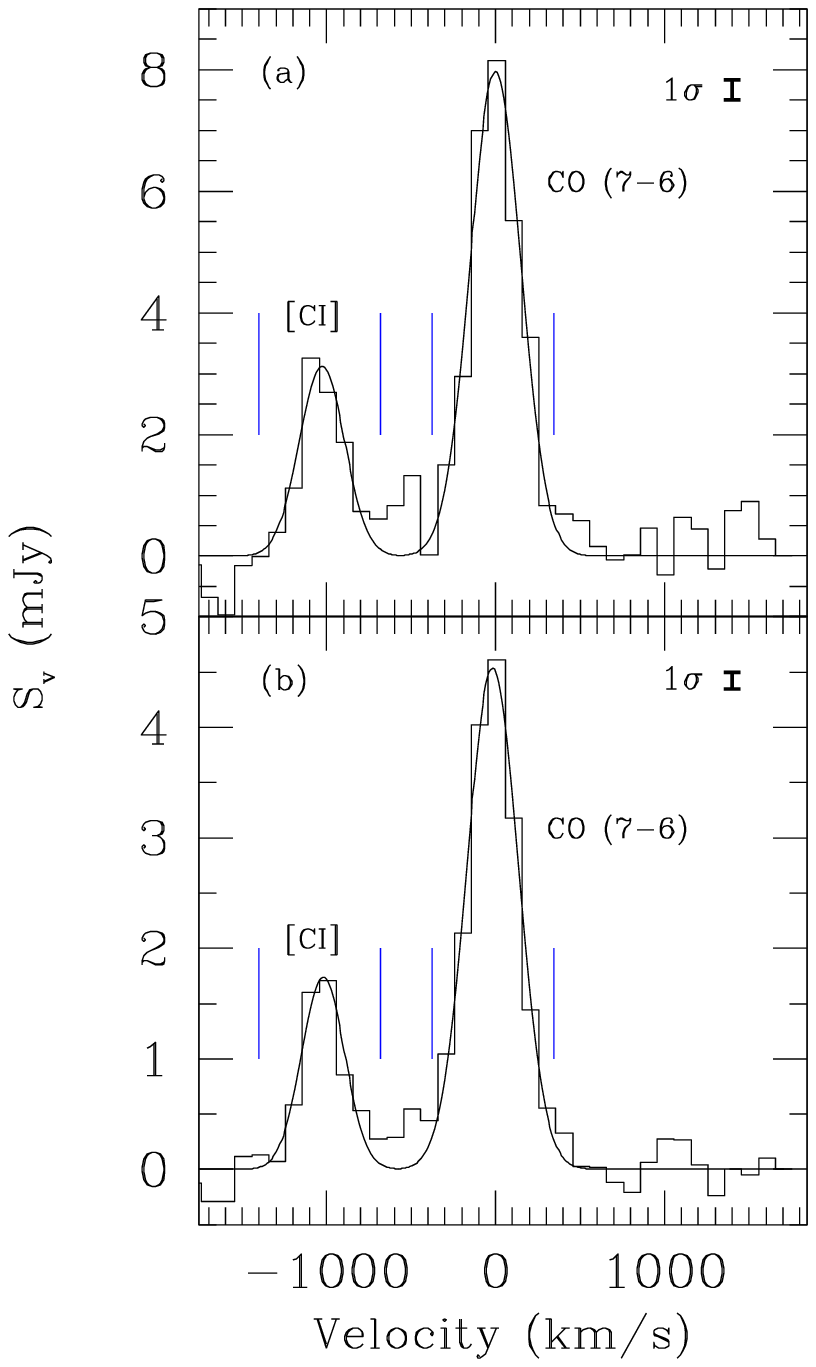} & \includegraphics[width=0.40\textwidth, bb=200 250 410 718]{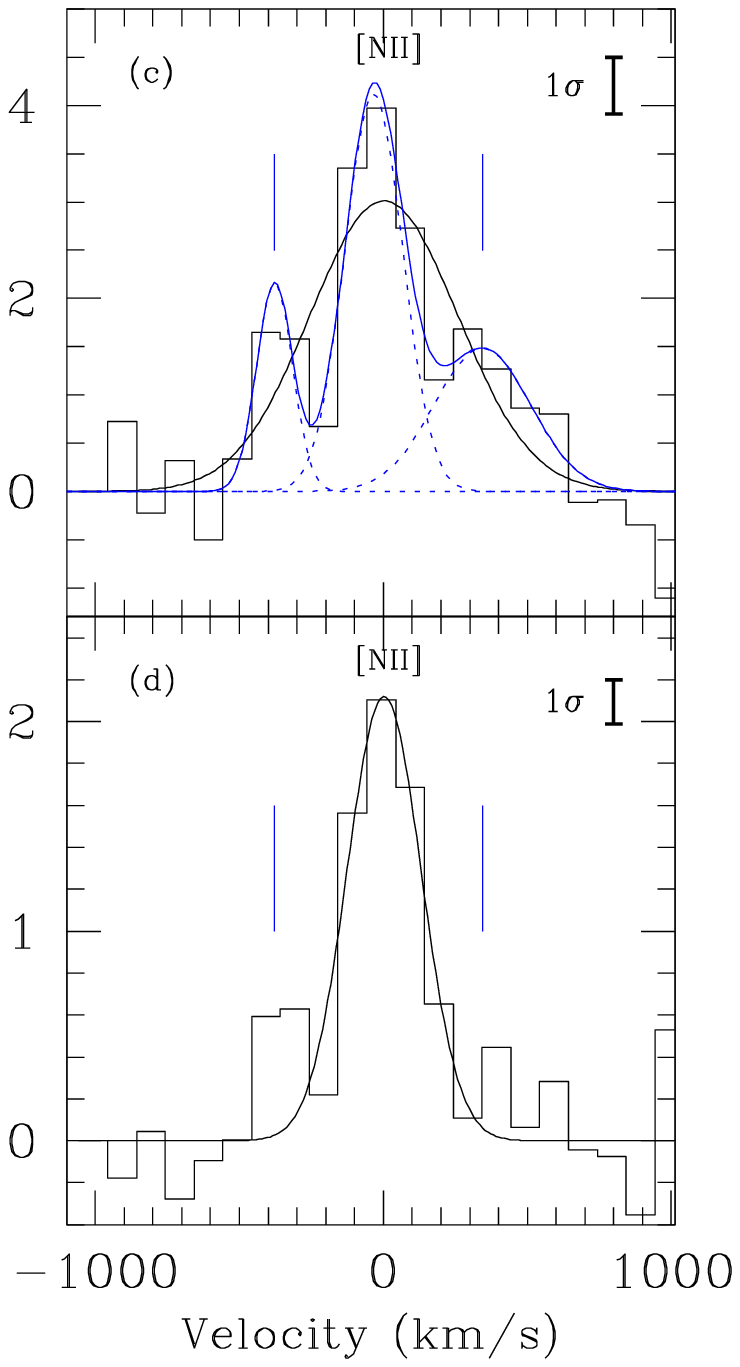} \\
\end{tabular}
\end{center}
\vspace{-0.8in}
\caption{
Extracted 1d spectra from the CO\,(7$-$6) (left panels) and \NII\ (right panels) 
spectral cubes, respectively, using an elliptical aperture of the size index $n
 = 2.5$ (upper panels) or $n = 1.0$ (lower panels). The aperture size index $n$ 
(see the text) is the common stretch factor for the major and minor axes of 
the aperture relative to the reference ellipse defined by the 2d Gaussian fit 
to the frequency-integrated line image.  The velocity scales are with respect to 
the CO\,(7$-$6) and \NII\ line central frequencies (in Table 1), respectively. 
Therefore, the observed central frequency of the \CI\ line is located at $-$1,025\kms.
The black solid curve in each panel is the best fit to the spectrum using either 
one or two Gaussian functions assuming zero residual continuum.   
The blue solid curve in panel (c) is a 3-component 
Gaussian fit to the spectrum with the individual components shown by the dotted 
curves in blue.  For each spectral line, a pair of blue vertical bars mark 
the velocities of the peaks of 
the two minor Gaussian components of the \NII\ emission in panel (c). The estimated 
1$\sigma$ error bar is shown in each panel.
}
\label{Fig3}
\end{figure*}

\begin{figure*}
\vspace{-0.5in}
\centering
\includegraphics[width=1.0\textwidth, bb=0 0 958 1008]{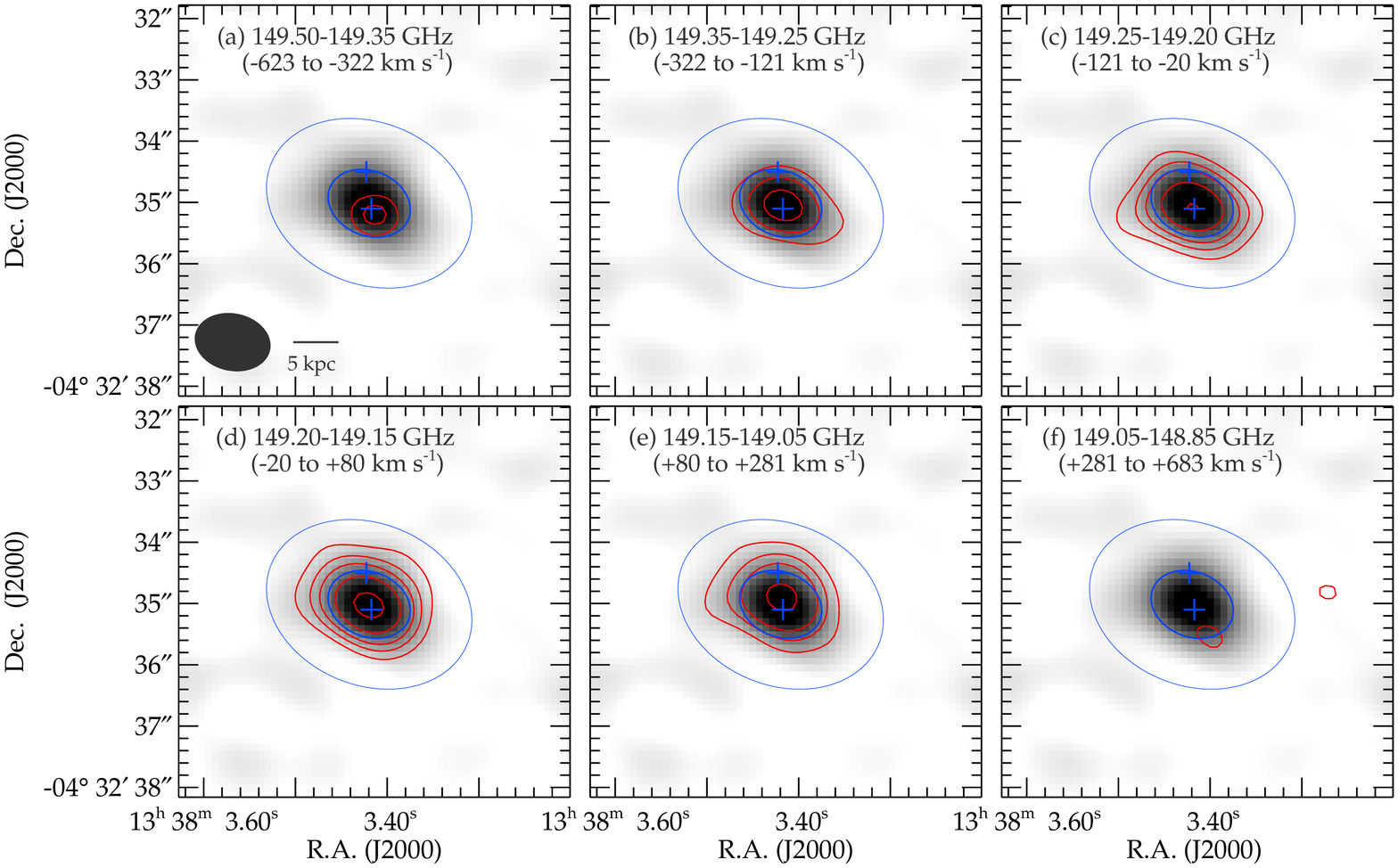}
\vspace{-2.7in}
\caption{
Velocity channel images of the CO\,(7$-6$) line emission with the channel 
frequency and velocity ranges noted in each panel. The gray image is always 
the total CO\,(7$-$6) image from Fig. 1g.  Starting at 
3 $\times$ the r.m.s. noise and being in units of Jy\,beam$^{-1}$\,\kms, 
the contour levels are: (a) [3.0, 3.6] $\times$ 0.052,
(b) [3, 6, 9] $\times$ 0.046, (c) [3, 6, 9, 15, 22] $\times$ 0.027, 
(d) [3, 6, 9, 15, 22] $\times$ 0.028, (e) [3, 6, 9, 15] $\times$ 0.041, 
and (f) [3.0] $\times$ 0.049.  The other over-plotted symbols are 
the same as in Fig. 1.
}
\label{Fig4}
\end{figure*}

\section{Analysis and Discussion} \label{sec3}

\subsection{{\rm CO\,(7$-$6)} Emission} \label{sec3.1}

The FWHM major axis from our 2d Gaussian fit to the CO\,(7$-$6)
image in Fig.~1g is 1.34\arcsec\,($\pm$0.06\arcsec) (at PA 
= 67\arcdeg$\pm$5\arcdeg), suggesting that the emission 
is barely resolved at best.  A 2d deconvolution of the source size 
with the ALMA beam (using the CASA function ``deconvolvefrombeam'')
results in a beam-deconvolved diameter of 
$d_{\rm deconv} \approx 0.6$\arcsec$\pm$0.1\arcsec 
(equivalent to 4.1$\pm$0.7\,kpc) at PA$_{\rm deconv} \approx
44$\arcdeg.
Fig. 1d shows that the peak surface brightness of the CO\,(7$-$6)
emission is close to the location of source S, suggesting 
that this source dominates the CO\,(7$-$6) flux.

The peak velocity dispersion of the CO\,(7$-$6) emission is 
$\sim$180\kms\ in Fig.~2e.  If one extracts a series of 
CO\,(7$-$6) spectra using apertures of the size index $n$
from 0.5 to 3.0, and fit a Gaussian profile to 
each spectrum extracted, the resulting line widths show
little change as the aperture 
size increases (see Table 2).  This suggests that most 
of the warm/dense molecular gas should be concentrated 
within an area smaller than the ALMA beam size.

Given the relatively high S/N obtained for the CO\,(7$-$6)
emission, we show in Fig.~4 six contiguous velocity channel
images for this line.  Fig. 4a is from the frequency range
where there is an excess emission between the main \CI\ and 
CO\,(7$-$6) profiles in Fig. 3a or 3b.  The peak emission
in Fig. 4a is significant at $\sim$3.6$\sigma$ and appears 
to be spatially associated with source S.  This signal 
is also clearly detected at a peak S/N $\approx$ 3.9 (3.5)
in Fig.~3a (3b) after we remove the two Gaussian line profiles
shown. The spectral signal peaks at a velocity of about 
$-500$\kms\ with respect to the line center of the CO\,(7$-$6)
emission. Its velocity width is rather uncertain, but appears
to be between 200 and 400\kms.   This emission could be 
either from (i) blue-shifted CO\,(7$-$6) emission or 
(ii) red-shifted \CI\ emission, at a maximum velocity of 
$\sim$600\kms\ if only one of the lines is the dominant 
contributor.  Another possibility is that (iii) blue-shifted 
CO\,(7$-$6) and red-shifted \CI\ emission are both present. 
In this case, each emission could have a lower maximum 
velocity.  (ii) is less likely because the redshifted 
\CI\ signal would amount to as much as 1/3 of the main line
peak. Therefore, there should be significant blue-shifted 
CO\,(7$-$6) emission regardless of whether (i) or (iii) 
is the actual case.  For the analyses that follow, we just 
assume that (i) is the case for the sake of simplicity.

The individual channel images in Fig. 4 also show that 
the inner emission contours center near source S at 
all velocities, especially below 80\kms.  
This suggests that the source S is indeed the dominant 
CO\,(7$-$6) emitter.  Fig. 4f shows the most red-shifted 
CO\,(7$-$6) emission detected at S/N $\gtrsim 3$. 
Corresponding to the emission excess seen on the red wing 
of the main CO\,(7$-$6) profile in Fig. 3a or 3b, this 
redshifted signal peaks spatially at a location only slightly 
south-west of the position of the most blue-shifted 
CO\,(7$-$6) emission in Fig. 4a.  Note that any 
CO\,(2$-$1) emission associated with the CO\,(7$-$6) line
emission in either Fig. 4a or 4b would be outside 
the bandwidth of Riechers et al. and therefore is not 
expected to be seen in their CO\,(2$-$1) observations.

The observed spatial locations of the blue- and red-shifted 
CO\,(7$-$6) emission is suggestive of a possible bipolar gas 
outflow.  The maximum velocity of the blue-shifted outflow 
could be as high as $\sim$600\kms\ although this estimate 
may be subject to contamination from any redshifted \CI\ 
emission (see Fig. 3a). On the red side, 
the maximum velocity of the outflow could be as high as 
$\sim$500\kms, subject to a considerable uncertainty 
associated with the relatively low S/N ratios achieved 
(see Fig. 3a). Even though gas outflows at velocities ranging 
from 250 to as high as 1,400\kms\ have been claimed in a number of 
high-$z$ QSOs (e.g., Wei\ss\ et al. 2012;  Maiolino et al. 2012;
Cicone et al. 2015; Feruglio et al. 2017),  the confirmation
of this possible CO\,(7$-$6) gas outflow in BRI 1335-0415 
still requires more sensitive observations in the future. 
Nevertheless, such an outflow should have a minor contribution
to the total flux of the CO\,(7$-$6) emission (see Fig. 3a) 
and a negligible effect on the line width of the Gaussian 
fit to the main emission profile (see Fig. 3b).

The CO\,(7$-$6) line width given in Table 1 is significantly 
narrower than, but still consistent (within about 1$\sigma$)
with the CO\,(5$-$4) line width of 420 $\pm$ 60\kms\ given
in Guilloteau et al. (1997).  However, their spectrum is  
at significantly lower S/N and has a narrower frequency 
coverage ($\sim$1,100\kms) than ours. 
Using their CO\,(5$-$4) line flux,  we derived that 
the CO\,(7$-$6)/CO\,(5$-$4) luminosity ratio is $\sim$1.3.  
This ratio suggests a rest-frame $C(60/100) > 1$  based on 
the template CO SLEDs in Lu et al. (2017a; see their Table 6).

The total CO\,(7$-$6) flux in Table 1 is 3.08\,($\pm 0.11$) Jy\,\kms\  
based on the Gaussian fit to the line profile in Fig. 3a. 
The total CO\,(2$-$1) flux from a VLA D-array observation 
is 0.62\,($\pm 0.03$) Jy\,\kms\ (Jones et al. 2016). This 
flux likely represents the total CO\,(2$-$1) flux better than
the flux of 0.43\,Jy\,\kms\ from the higher-resolution 
observation in Reichers et al. (2008)  This translates to 
a line luminosity ratio of $L_{\rm CO(7-6)}/L_{\rm CO(2-1)} 
\approx 17.4$.  In comparison, the CO\,(7$-$6) and CO\,(2$-$1)
fluxes of Arp 220 are $9.29\,(\pm 0.19)$ and $1.27\,(\pm 0.04)\,
\times 10^{-17}$\,W\,m$^{-2}$, respectively (Lu et al. 2017a; 
Mart\'in et al. 2011), resulting in $L_{\rm CO(7-6)}/L_{\rm CO(2-1)} 
\approx 7.3$.  Therefore, the warm CO emission in BRI 1335-0417
is twice as prominent as that in Arp 220, one of the brightest 
local (U)LIRGs.

\subsection{{\rm \CI\ 370\um}\ Emission} \label{sec3.2}

Like the CO\,(7$-$6) emission, the \CI\ emission in Fig. 1h
is also barely resolved at best, with $d_{\rm deconv} = 
0.7$\arcsec\,($\pm$0.2\arcsec) (equivalent to 5$\pm$1 kpc) at 
PA$_{\rm deconv} \approx 42$\arcdeg.  This $d_{\rm deconv}$ 
is apparently larger than that for the CO\,(7$-$6) emission
although the difference is not statistically significant.
Our estimated 
$d_{\rm deconv}$ value of the \CI\ emission is comparable to 
the scale of the CO\,(2$-$1) distribution seen in the {0.23\arcsec}-resolution 
VLA map in Riechers et al.~(2008).  This is not unexpected as 
the \CI\ emission in local galaxies scales not only in flux 
(e.g., Jiao et al. 2017; Lu et al. 2017a), but also in spatial
distribution with a low-$J$ CO line emission such as CO\,(1$-$0)
or CO\,(2$-$1)  (e.g., Ojha et al. 2001; Ikeda et al. 2002; 
Beuther et al. 2014).

The extracted 1d spectrum (Fig. 3a) of the \CI\ emission
has a peak flux density of 3 mJy (see Fig. 3a), consistent
with the upper flux limit of 5.1 mJy set for this line 
in Walter et al. (2011).   Table 2 shows that the \CI\ line 
velocity width increases moderately with
an increasing aperture used for the spectrum extraction. 
For example, the line FWHM increases by 15\% from 293 to 
338\kms\ when the aperture size index increases from $n 
= 0.5$ to $n = 3.0$.  Even though the amount of the increase
is statistically significant only at 2-3$\sigma$,  
the increase is systematic and more noticeable than that 
in the CO\,(7$-6$) line.  This suggests that the spatial
distribution of the molecular gas traced by the \CI\ line 
is effectively broader than the denser molecular gas
traced by the CO\,(7$-$6) line.  Furthermore, for the aperture
of $n \le 1$, the FWHM of the \CI\ line is meaningfully 
(i.e., at $\gtrsim$4$\sigma$) less than that of 
the CO\,(7$-$6) line.  This line width difference in 
the inner region of the galaxy is consistent with 
the velocity dispersion difference between these lines 
in Fig. 2.  Therefore, the gas traced by the \CI\ emission 
is effectively less concentrated towards the galaxy center 
than the warm dense gas traced by the CO\,(7$-$6) emission. 
Likewise, the ionized gas dominating the \NII\ emission 
distributes over a much larger spatial scale than the molecular 
gas dominating the \CI\ emission.

\begin{deluxetable*}{lccccc}
\tabletypesize{\small}
\tabletypesize{\footnotesize}
\tabletypesize{\scriptsize}
\tablenum{2}
\tablewidth{0pt}
\tablecaption{Gaussian Line Profile: Central Frequency and FWHM$^a$}
\tablehead{
\colhead{Line}      & \colhead{$n = 0.5$}   & \colhead{$n = 1.0$}                  & \colhead{$n = 2.0$} & \colhead{$n = 2.5$}                & \colhead{$n = 3.0$}  \\
\colhead{(1)} 	    & \colhead{(2)}         & \colhead{(3)}                        & \colhead{(4)}       & \colhead{(5)}                      & \colhead{(6)}}
\ \ \ CO\,(7$-$6)   &   (149.189, 360)      &  (149.188$\pm 0.002$, 360$\pm \ 9$)    &  (149.181, 359)     & (149.179$\pm 0.003$, 354$\pm 11$)  & (149.179, 350) \\
\ \ \ \CI\          &   (149.685, 293)      &  (149.687$\pm 0.003$, 298$\pm 13$)   &  (149.688, 303)     & (149.690$\pm 0.004$, 314$\pm 16$)  & (149.692, 338) \\
\ \ \ \NII\         &   (270.236, 280)      &  (270.237$\pm 0.011$, 302$\pm 26$)   &  (270.241, 471)     & (270.238$\pm 0.028$, 603$\pm 64$)  & (270.237, 694) \\
\enddata
\tablenotetext{a}{$n =$ the aperture size index. Each pair of numbers in brackets are the line central frequency in GHz and 
the line FWHM in \kms, both from a 1d Gaussian line profile fitting.  For simplicity, the uncertainties are shown only for 
the cases of $n=1.0$ and 2.5.
}
\end{deluxetable*}

The observed \CI\ flux in Table 1 corresponds to $L' = 
(1.6 \pm 0.1) \times 10^{10}$\,K\,\kms\,pc$^2$.  Using 
a sample of local (U)LIRGs with both CO\,(1$-$0) and 
\CI\ flux measurements, Jiao et al. (2017) showed that 
there is a nearly linear correlation between the logarithmic
line luminosities of the CO\,(1$-$0) and \CI\ emission, 
with a scatter of $\sim$0.19 dex. This observation is 
the basis for their derivation of a mean relationship 
between the total molecular gas mass, $M_{\rm gas}$, and 
the \CI\ line luminosity.   In practice, since the \CI\ 
line has a modest $T_{\rm ex} \approx 63\,$K, its 
luminosity has a dependence on the gas temperature.
To some extent, this systematic effect has been accounted
for by the observed scatter around their mean correlation 
between $M_{\rm gas}$ and the \CI\ luminosity, which 
translates to their claimed uncertainty of a factor of 
2$-$3 in the derived $M_{\rm gas}$.   Using their eq. (11), 
we derived $M_{\rm gas} \approx 4.4 \times 10^{11}\,M_{\odot}$.  
This gas mass estimate is only 5 times larger than 
the molecular gas mass of 9.2 $\times 10^{10}\,M_{\odot}$
estimated by Riechers et al. (2008) based on their 
CO\,(2$-$1) flux and the typical ULIRG CO-to-H$_2$ mass
conversion factor.  The difference could be easily accounted
for by the differences between the gas mass estimators 
used (see Jiao et al. 2017) and the fact that our observation
used a larger synthesized beam.

For local (U)LIRGs, the ratio $L_{\rm [CI]}/L_{\rm CO(7-6)}$ 
decreases significantly as $C(60/100)$ increases (Lu et al. 2017a).  
Fig. A1 in the appendix illustrates this correlation using 
a plot reproduced from Lu et al. (2017a). For BRI~1335-0417, 
the observed $\log\,L_{\rm [CI]}/L_{\rm CO(7-6)} = -0.47$ 
($\pm 0.04$), which corresponds to $C(60/100) \sim 1.21$ 
($\pm 0.11$) based on the solid line in Fig. A1.  In comparison,
$\log L_{\rm [CI]}/L_{\rm CO(7-6)} \approx -0.30$ for Arp 220.

Since $L_{\rm CO(7-6)} \propto$ SFR and $L_{\rm [CI]} \propto
L_{\rm CO(1-0)}$, the correlation in Fig. A1 suggests that 
the ratio SFR/$L_{\rm CO(1-0)}$ increases by a factor of $\sim$6
as $C(60/100)$ increases from $\sim$0.4, which is not too 
different from the FIR colors for local normal, star-forming
galaxies, to $1.2$, where most ULIRGs tend to show up. 
This magnitude of the variation in SFR/$L_{\rm CO(1-0)}$ is 
roughly what is seen in Genzel et al. (2010; see their Fig. 2). 
Fig.~A1 shows that the change in SFR/$L_{\rm CO(1-0)}$ is not 
bi-modal, but a continuous function of $C(60/100)$ (see also
Gao \& Solomon 2004; Cheng et al. 2018).  If $L_{\rm [CI]}$ 
is proportional to $M_{\rm gas}$, the ratio $L_{\rm CO(7-6)}/L_{\rm [CI]}$ 
merely measures the so-called SF efficiency (SFE).  

\subsection{{\rm \NII\ 205\um} Emission} \label{sec3.3}

The \NII\ emission appears to be more extended than 
both CO\,(7$-$6) and \CI\ emission, and has a FWHM 
major axis of 1.7\arcsec\,($\pm$0.3\arcsec) (see Table 1).
This leads to $d_{\rm deconv}$ = 1.3\arcsec$\pm$0.3\arcsec\ 
(or 9$\pm$2 kpc) at PA$_{\rm deconv} \approx 44$\arcdeg.  
This size is about twice as large as that for the CO\,(7$-$6)
emission, with the observed size difference at a significance
level of $\sim$2.3$\sigma$. 
The phenomenon that the \NII\ emission is much more 
extended than both of the underlying dust continuum 
and CO\,(7$-$6) emission is also seen in some 
other high-$z$ galaxies.  For example, the \NII\ emission
is at least twice as extended as the dust continuum 
emission for both the QSO and SMG in the compact, 
interacting galaxy group BR~1202-0725 at $z = 4.7$ 
(Lu et al. 2017b).

It is interesting to notice that the major-axis PA's 
of all three emission lines, after the deconvolution 
with the ALMA beam, are around 44\arcdeg.  The PA from 
the \NII\ emission should be reliable as the line emission
is well resolved spatially. 
This deconvolved PA is not in the direction along sources 
S and N, but is more aligned with the major axis of 
source S.  This may simply reflect the fact that source S 
is the dominant one in the system.

The extracted 1d spectrum of the \NII\ emission in Fig. 3c 
appears to comprise a core component and two fainter, but
distinct wing components, albeit with both of them having 
a peak S/N just below 3.    While the brightest part
of the blue-wing image is near source S (see Fig. 1c), 
that of the red-wing image is in proximity of source N 
(see Fig. 1d).  This pattern, along with the fact that both 
wings appear as a long, narrow structure over a projected 
scale of 15-20 kpc, is reminiscent of tidal gas 
tails in a galaxy major merger, rather than related to 
the possible CO\,(7$-$6) outflows discussed in \S3.1, which 
appear to largely align with the line of sight.  However, 
these \NII\ wing components are only detected at S/N $\sim$ 
2$-$3,  the tidal tail picture remains as a speculation 
at this point.

We also fit 3 Gaussians to the full \NII\ spectrum  and the best  
fit is shown by the solid blue curve in Fig. 3c, with 
the dotted blue curves representing the individual Gaussian 
components. These Gaussian fits are also given in Table 1. 
The peaks of the two minor Gaussian components correspond respectively
to the velocity shifts of -355 and +340 \kms\ relative to the main 
Gaussian peak.  In Fig. 3. these velocities are marked by the two 
blue bars for each of the 3 lines.   We note that the 
reduced Chi-squared values are 0.64 and 1.29 for the 3-Gaussian
and single-Gaussian fits in Fig.~3c, respectively, indicating 
that the single Gausssian is statistically a better fit although
both fits are not very satisfactory. We also tried a couple of 
ways to fit the data with 2 Gaussian components. The resulting 
reduced Chi-squared values of $\sim$1.34 are slightly larger 
than that for the single Gaussian fit.  We therefore 
derived two \NII\ luminosities
in Table 1 from Fig. 3c. One is based on the flux from the single 
Gaussian fit to the \NII\ spectrum and the other based on the flux 
of the central Gaussian component in the 3-Gaussian component fit.  
We use these respectively as the upper and lower limits on the true 
\NII\ luminosity in our line luminosity ratio analysis below.

\begin{figure}
\vspace{-0.3in}
\centering
\includegraphics[width=0.55\textwidth, bb=50 144 592 718]{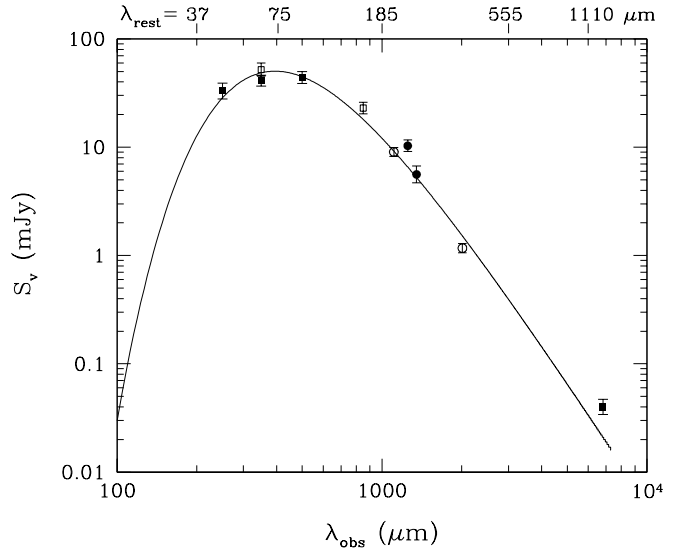}
\vspace{-1.0in}
\caption{Plot of continuum flux measurements as a function of the observed wavelength.
(The corresponding range of the rest-frame wavelength is indicated on the top.)
Different symbols indicate the sources of the data: filled squares (Wagg et al. 2014),
open squares (Benford et al. 1999), open circles (this work), and filled circles 
(Guilloteau et al. 1997).  The solid curve is the best graybody SED fit (with a fixed
power-law emissivity index $\beta = 1.8$) to the data based on an equal weight basis, 
yielding $T_{\rm dust} = 41.5$\,K.  
}
\label{Fig5}
\end{figure}

\subsection{Line Luminosity Ratios and FIR Dust Emission} \label{sec3.4}

The continuum fluxes at $\lambda_{\rm obs} \approx 1108$ and 
$2011$\um\ measured in this work are shown in Fig. 5, along 
with other published fluxes at various wavelengths. 
The solid curve in Fig.~5 is the best gray-body 
spectral energy distribution (SED) fit to all the data points 
with a fixed dust emissivity power
law index $\beta = 1.8$ (Planck Collaboration et al. 2011).  
This SED fit gives $T_{\rm dust} = 41.5\,$K or $C(60/100) \approx
1.1$.

We also used various line luminosity ratios to estimate the value
of $C(60/100)$ using the luminosity values calculated in Table 1.  
The resulting estimates of $C(60/100)$ 
are summarized in Table 3 and range from $\sim$1.05 to $\sim$1.20, 
with a typical uncertainty of 0.10 or 0.15 in each estimate.  
The average value of 1.1 ($\pm$0.1) for $C(60/100)$ corresponds
to $T_{\rm dust} \approx 41$\,($\pm$2)\,K if $\beta$ is also fixed 
at 1.8.  This temperature estimate is in good agreement with that 
from the direct SED fit.

The FIR (42$-$122\um) luminosity, $L_{\rm FIR}$, is 1.9 $\times 
10^{13}\,L_{\odot}$ from the SED fit in Fig.~5, resulting in 
$\log L_{\rm CO(7-6)}/L_{\rm FIR} \approx -4.46$.  This value is 
within $\sim$1.2$\sigma_s$ of the average value of this ratio
for the local SF-dominated LIRGs in Lu et al. (2015),  where 
$\sigma_s$ is the local sample standard deviation. This result
confirms that the CO\,(7$-$6) emission in the QSO BRI 1335-0417 
is an excellent SFR predictor and that the FIR luminosity is also
dominated by the SF in this QSO.

\begin{deluxetable}{lllc}
\tabletypesize{\scriptsize}
\tablenum{3}
\tablewidth{0pt}
\tablecaption{Line Luminosity Ratio}
\tablehead{
\colhead{Line luminosity ratio} 		  & \colhead{Value}          & \colhead{$C(60/100)^a$}  & \colhead{Ref$^b$} \\
\colhead{(1)} 		                     & \colhead{(2)}             & \colhead{(3)}            & \colhead{(4)}}
$\log$\,(\NII$_{\rm core}$/CO\,(7$-$6))         & $-$0.21\,($\pm$0.06)$^c$    &	    1.20 ($\pm$0.15)     &   (1) \\
$\log$\,(\NII$_{\rm total}$/CO\,(7$-$6))        & $+$0.06\,($\pm$0.06)$^c$    &	    1.06 ($\pm$0.15)     &   (1) \\
$\log$\,(\CII/CO\,(7$-$6))		        & $+$1.30\,($\pm$0.09)$^c$    &	    1.04 ($\pm$0.15)   	 &   (1) \\  
$\log$\,(\CI/CO\,(7$-$6))			        & $-$0.47($\pm$0.04)$^d$    &	    1.20 ($\pm$0.11)      &   (2,3) \\
CO\,(7$-$6)/CO\,(5$-$4)$^e$                   & $+$1.30\,($\pm$0.15)$^c$    &	    $>$1.0   		 &   (2) \\
\enddata
\tablenotetext{a}{The uncertainty is dominated by the scatter of the line ratios of the local calibration 
		  sample, i.e., 0.1 to 0.15.}
\tablenotetext{b}{References for converting a line luminosity ratio to $C(60/100)$: (1) Lu et al. (2015); 
(2) Lu et al. (2017a); (3) Appendix of this work.}
\tablenotetext{c}{The uncertainty calculated assumes a total flux uncertainty of 10\% for the line flux measured in this work.}
\tablenotetext{d}{The uncertainty given here is free from any ALMA systematic flux calibration error.}
\tablenotetext{e}{The CO\,(5$-$4) flux is taken from Guilloteau et al. (1997).} 
\end{deluxetable}

\begin{figure*}
\centering
\includegraphics[width=1.0\textwidth, bb=18 144 592 680]{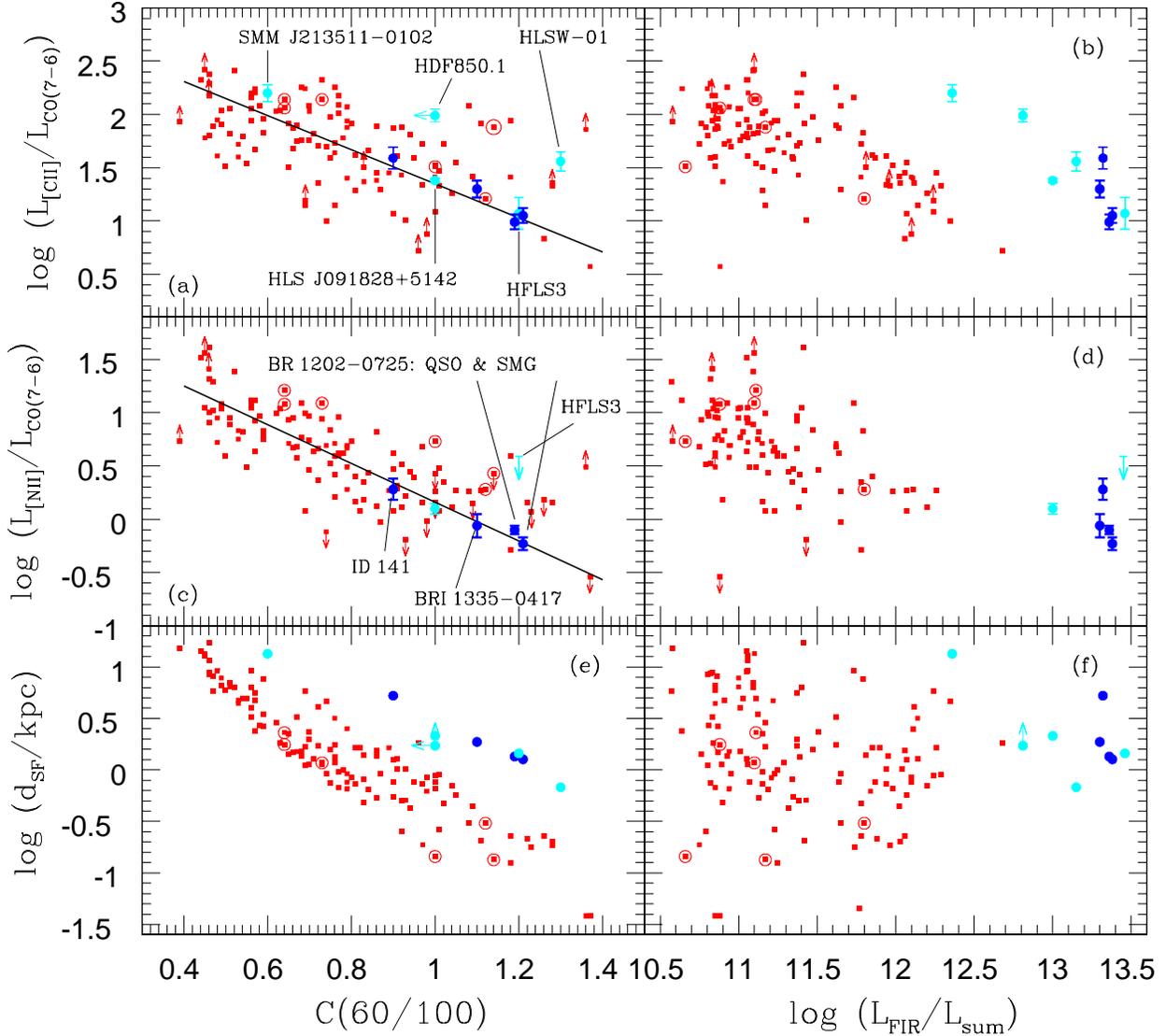}
\vspace{-0.7in}
\caption{Plots of $\log L_{\rm [CII]}/L_{\rm CO(7-6)}$ (top panels),
$\log L_{\rm [NII]}/L_{\rm CO(7-6)}$ (mid panels) and 
$\log d_{\rm SF}$ (bottom panels) as a function
of the FIR color (left panels) or $\log\,L_{\rm FIR}$ (right panels), 
for the local (U)LIRG sample (filled squares in red) and a number of 
distant ULIRGs at $2 < z < 6.5$ (filled circles in cyan or blue) from 
Lu et al. (2015), which also has an explanation for the labelling of 
the high-$z$ galaxies here. 
The data points further circled in red are the local (U)LIRGs, for 
which the AGN contribution is likely to dominate the bolometric 
luminosity based on a number of mid-IR diagnostics (see Lu et 
al. 2015).  Those high-$z$ targets that are also in our ALMA program 
are plotted in blue (Lu et al. 2017b and this work), including ID~141
for which we used $C(60/100)$ and $L_{\rm FIR}$ from Cox et al.~(2011),
a gravitational de-magnification factor of 4.1 from Brussmann et al.~(2012)
and our own \NII\ line flux measurement in Cheng et al. (2018, in preparation). 
The solid lines in (a) and (c) represent the average correlation of 
the local sample given in Lu et al. (2017b). The value of $d_{\rm SF}$ 
is derived from the $L_{\rm FIR}$-based SFR and $C(60/100)$ as 
prescribed in Lu et al. (2015) and has an uncertainty of about a 
factor of 2.
}
\label{Fig6}
\end{figure*}

\subsection{Star Formation and Gas Properties} \label{sec3.5}

Following Lu et al. (2015), the SFR inferred from $L_{\rm CO(7-6)}$
is 5.1 $\times 10^3$\,$M_{\odot}\,$yr$^{-1}$ using the initial mass 
function (IMF) of Chabrier (2003).   For local (U)LIRGs, 
$\Sigma_{\rm SFR}$ is empirically correlated with the rest-frame 
$C(60/100)$ (Liu et al. 2015) or $f_{\nu}(70\um)/f_{\nu}(160\um)$ 
(Lutz et al. 2016).   The scatter of these correlations is still 
fairly significant,  e.g., $\sim$0.6 dex in Lutz et al. (2016).  
Nevertheless,  these two independent correlations, together with
the flux densities from our SED fit, give comparable estimates of 
$\Sigma_{\rm SFR} \sim 1 \times 10^3\,M_{\odot}$\,yr$^{-1}$\,kpc$^{-2}$ 
after adjusting them to Chabrier IMF and 
increasing the \LFIR-based SFR in Lutz \etal by a factor of 
2 to align with the Kennicutt (1998) formula.  These estimates of
$\Sigma_{\rm SFR}$ are quite high, approaching the Eddington
limit on the order of $10^3\,M_{\odot}$\,yr$^{-1}$\,kpc$^{-2}$ 
(Murray et al. 2005; Thompson et al. 2005; Hopkins et al. 2010).

The face-on FWHM diameter, $d_{\rm SF}$, of the SF region is 
estimated to be 1.7$^{+1.7}_{-0.8}$\,kpc, via $\Sigma_{\rm SFR}
= (\frac{1}{2}\,{\rm SFR})/({\frac{1}{4}}\pi\,d^2_{\rm SF})$.
While this estimated $d_{\rm SF}$ is smaller than 
the de-convolved diameter of $d_{\rm deconv} = 4.1\pm$0.7\,kpc
from the image analysis of the CO\,(7$-$6) emission,  
the difference is only significant at $\sim$1.3$\sigma$.
Following Scoville et al. (2016), we also derived $M_{\rm gas} 
\approx  5 \times 10^{11}\,M_{\odot}$ based on the rest-frame 
$f_{\nu}$(850\um) from our continuum SED fit.   The formal uncertainty 
for this $M_{\rm gas}$ estimate is about a factor of 2.  
This is in good agreement with the molecular gas mass inferred
from our \CI\ flux above. The characteristic 
gas depletion time $\tau_{\rm gas}$ ($\equiv M_{\rm gas}$/SFR) 
is $\sim$$10^8$\,years.

The derived SFR of $5.1 \times 10^3\,M_{\odot}\,$yr$^{-1}$ 
for BRI 1335-0417 is similar to that of the QSO and SMG in 
the BR~1202-0725 galaxy group system at $z = 4.7$ (Lu et al. 2017b), 
making BRI 1335-0417 one of the high-$z$ galaxies with the highest
SFR known.  Among the brightest ULIRGs in the local Universe, 
Arp 220 has $L_{\rm CO(7-6)} \approx 2.2 \times 
10^7\,L_{\odot}$ and $L_{\rm [CI]} 
\approx 1.1 \times 10^7\,L_{\odot}$ (Lu et al. 2017a).  Therefore,
the $L_{\rm CO(7-6)}$-based SFR and $L_{\rm [CI]}$-base $M_{\rm gas}$
ratios of BRI 1335-0417 to Arp 220 are $\sim$35 and $\sim$24, respectively.
The ratio of these two values is $\sim$1.5, which implies a 
50\% larger SFE for BRI 1335-0417. 

\subsection{On Merger-induced SF at High $z$} \label{sec3.6}

In Fig. 6 we compare BRI\,1335-0417 and a few other high-$z$ 
ULIRGs with the local (U)LIRG sample in Lu et al. (2015) via 
plots,  all as a function of $C(60/100)$ (left panels) or 
$\log\,L_{\rm FIR}$ (right panels), of $\log\,L_{\rm [CII]}/L_{\rm CO(7-6)}$ 
(two top panels), $\log\,L_{\rm [NII]}/L_{\rm CO(7-6)}$ (mid panels) and
$\log\,d_{\rm SF}$ (bottom panels).   The $d_{\rm SF}$ values
are derived from the $L_{\rm FIR}$-based SFR and $C(60/100)$-based 
$\Sigma_{\rm SFR}$ as prescibed in Lu et al. (2015). That is,
$\log\,d_{\rm SF} = 0.5\,\log\,L_{\rm FIR} - 5.04\,\log\,C(60/100)
- 6.17$.  The error bars for these $d_{\rm SF}$ estimates are on 
the order of a factor of 2.   All high-$z$ galaxies 
plotted here have a $C(60/100)$
derived from the FIR dust SED fitting and their $L_{\rm FIR}$ 
corrected for a gravitational lensing magnification factor 
when appropriate (see Carilli \& Walter 2013).

Despite the fact that some of the high-$z$ galaxies in Fig. 6
are an order of magnitude brighter in $L_{\rm FIR}$ than 
the brightest local ULIRGs, the local and high-$z$ galaxies
appear to follow the same average correlation in each of 
the two top-left panels in Fig. 6. On the other hand, clear 
segregations are seen between the local and high-$z$ galaxies
in the two top-right panels in Fig. 6.  This suggests clearly 
that it is $C(60/100)$ or $\Sigma_{\rm SFR}$, {\it rather than}
$L_{\rm FIR}$, that drives these line flux ratios.
Therefore, a plot of $\log\,L_{\rm [CII]}/L_{\rm FIR}$ (where 
$L_{\rm FIR}$ can be substituted by $L_{\rm CO(7-6)}$) against
$L_{\rm FIR}$ offers diminished diagnostic value in spite 
of the fact that such plots have been widely used in the literature
for galaxy samples across different redshift epochs.

Furthermore, the two bottom panels in Fig. 6 reveal that,
with SFR surface densities comparable to that of local ULIRGs, 
high-$z$ hyperluminous infrared galaxies, such as BRI 1335-0417, 
reach their higher global SFR mainly via a larger star-forming
area.  This is in an agreement with 
a similar conclusion reached by Rujopakarn et al. (2011) who
estimated  $\Sigma_{\rm SFR}$ based on a radio continuum size
in galaxies up to $z \sim 2.5$.  The phenomenon that the high-$z$ 
luminous galaxies  have a larger SF area when compared with 
local ULIRGs is one of the motivations behind the so-called cold 
gas accretion scenario as the main SF mode for most of the SMGs 
at high $z$ (e.g., Agertz et al. 2009; Dekel et al. 2009; 
Dav\'e et al. 2010), as opposed to the merger-triggered SF 
scenario (e.g., Tacconi et al. 2006, 2008).  
In the case of BRI 1335-0417, our observational results favor 
a merger-induced SF scenario.  These include (a) the very 
warm FIR color of $C(60/100) > 1$, (b) the CO\,(7$-$6) emission
is spatially more compact than the \CI\ emission, and 
(c) a significantly different velocity dispersion between 
the CO\,(7$-$6) and \CI\ in the inner region of the galaxy.
Together, they suggest that the multiple gas phases traced
by these lines arise effectively from different physical 
regions. Taking into consideration the much larger spatial 
scale of the \NII\ emission observed, our results appear to 
agree with the predictions from cosmological simulations
on galaxy mergers (e.g., Sparrie \& Springel 2016).

It is possible that many of the other high-$z$ galaxies 
with a warm FIR color could also be mergers, e.g., both 
the SMG and QSO in the BR~1202-0725 system (Lu et al. 2017b) 
also plotted in Fig. 6.  These galaxy systems might be 
caught near the maximum SFR phase
of a non head-on major galaxy merger, which starts 
before the two progenitors coalesce (e.g., Sparre \& 
Springel 2016). This could make the apparent SF size 
larger than that in a head-on galaxy merger in which 
the maximum SF phase occurs when the two galaxy nuclei 
coalesce.  Local examples might be the Antennae Galaxies
(Wang et al. 2004) or those widely-separated ULIRGs 
(Dinh-V-Trung et al. 2010).
However, the higher gas content in high-$z$ galaxies 
could lead to a merger-triggered $\Sigma_{\rm SFR}$ 
much higher than that in the Antennae system.  In fact, 
the inferred values of $\Sigma_{\rm SFR}$ for BRI 1335-0417 
and the two ULIRGs in BR~1202-0725 (Lu et al. 2017b) 
all approach the Eddington limit.  Detailed analyses of 
local ULIRGs also indicate that some of the advanced 
mergers in the local Universe also have a $\Sigma_{\rm SFR}$ 
near the Eddington
limit, but with a much smaller SF area (Barcos-Mu\~noz 
et al. 2017).  This might naturally explain the comparable
$C(60/100)$ (or $\Sigma_{\rm SFR}$) values in both 
local ULIRGs and many of the high-$z$ galaxies in 
Fig. 6.  In such a frame work, a head-on galaxy merger
would easily trigger a super Eddington $\Sigma_{\rm SFR}$
that could plausibly disrupt or end the starburst in 
a few million years, giving birth to a post-starburst
galaxy.  It would be interesting to explore whether 
the above conjecture may play a significant role in 
explaining the observations that there is already a 
population of post-starburst galaxies at $z > 4$ (e.g., 
Wiklind et al. 2008; Fontana et al. 2009;  Richard et
al. 2011; Straatman et al. 2014; Decarli et al. 2017).
It is not quite intuitive to see why cold gas accretion 
would stop in these quiescent galaxies, but not the other
SMGs at the same redshift epoch, under the orderly 
rotating disk SF scenario.

\section{Summary} \label{sec4}

We presented our recent ALMA observations of the CO\,(7$-$6), 
\CI\ 370\um\ and \NII\ 205\um\ line emission of BRI\,1335-0417, 
an infrared hyperluminous quasar at $z =$ 4.407, which is 
likely a major merger between two galaxies (N and S) based 
on a previous high-resolution CO\,(2$-$1) imaging.  Our main 
results can be summarized as follows:

At the achieved resolutions of $\sim$1.1\arcsec\ to 1.2\arcsec\ 
(or 7.5 to 8.2 kpc), the continuum emission at 205 and 372\um\ 
(rest-frame), the CO\,(7$-$6), and the \CI\ emission are all 
barely resolved at best.  On the other hand, the \NII\ emission 
is clearly resolved with a (FWHM) Gaussian major axis $d_{\rm 
deconv} = 9\pm$2 kpc after a deconvolution with the ALMA beam.

The observed CO\,(7$-$6) emission not only confirms that source
S dominates the SFR, but also reveals a complicated dynamics 
for the dense molecular gas.  This includes a possible bipolar 
outflow with a maximum velocity as high as 500 to 600\kms, 
likely driven by the QSO in source S.

While the three lines display similar large-scale velocity 
fields, they show different velocity dispersion fields: 
the CO\,(7$-$6) line displays a patchy velocity dispersion
field with a peak dispersion of $\sim$180\kms,  the \CI\
and \NII\ lines display smoother velocity dispersion fields, 
reaching a peak value of only $\sim$120 and $\sim$80\kms, 
respectively.  This suggests that the three lines arise 
effectively from different regions, a picture that fits
better with a merger-trigged SF scenario than with an orderly 
rotating disk SF scenario.

The far-infrared (FIR) dust temperature ($T_{\rm dust}$) of 
41.5\,K from the gray-body fit to the continuum measurements
agrees well with the average $T_{\rm dust}$ inferred from 
the line luminosity ratios of $L_{\rm [NII]}/L_{\rm CO(7-6)}$,
$L_{\rm [CII]}/L_{\rm CO(7-6)}$ and $L_{\rm [CI]}/L_{\rm CO(7-6)}$. 
The resulting $L_{\rm CO(7-6)}$/\LFIR\ is consistent with 
that of the SF-dominated luminous infrared galaxies in the local
Universe, thus confirming the validity of using $L_{\rm CO(7-6)}$
as a SFR tracer for such high-$z$ dusty QSOs.  We estimated 
a SFR of 5.1\,($\pm$1.5) $\times 10^3\,M_{\odot}$\,yr$^{-1}$
(assuming the Chabrier initial mass function),  an effective
diameter of 1.7$^{+1.7}_{-0.8}$\,kpc for the SF region, a near 
Eddington-limit SFR surface density of 
$\sim$$1\,\times\,10^3\,M_{\odot}$\,yr$^{-1}$\,kpc$^{-2}$, and 
a molecular gas mass of $\sim$$5\,\times\,10^{11}\,M_{\odot}$.

We show quantitative evidence that, with SFR surface densities 
comparable to that of local ULIRGs, high-$z$ hyperluminous 
infrared galaxies, such as BRI 1335-0417, reach their higher 
global SFR mainly via a larger star-forming area.

\acknowledgments

We thank an anonymous referee for a number of helpful comments.
This paper makes use of 
the following ALMA data: ADS/JAO.ALMA\#2015.1.00388.S. 
ALMA is a partnership of ESO (representing its member 
states), NSF (USA) and NINS (Japan), together with NRC 
(Canada), NSC and ASIAA (Taiwan), and KASI (Republic of 
Korea), in cooperation with the Republic of Chile. 
The Joint ALMA Observatory is operated by ESO, AUI/NRAO
and NAOJ.  This work is supported in part by the National 
Key R\&D Program of China grant \#2017YFA0402704, 
the NSFC grant \#11673028, and the Chinese Academy of 
Sciences (CAS), through a grant to the CAS South America 
Center for Astronomy  (CASSACA) in Santiago, Chile.
C.C. is supported by CAS through the CASSACA postdoc and 
visiting scholarship grant administered by the CASSACA, 
NAOC.


\vspace{0.5in}

\appendix

\section{\CI\ to CO\,(7$-$6) Luminosity Ratios of Local LIRGs}

Fig. A1 is a reproduction of Fig. 17c in Lu et al. (2017a).  We carried 
out least-squares fits to all the data points with both CO\,(7$-$6) and 
\CI\ lines detected, excluding the seven AGN-dominant LIRGs (circled 
in red) and NGC 6240 (circled in blue).  The least-squares bisector 
(Isobe et al.~1990) result, given in eq. (1), is shown by the solid line.  
The scatter around the fit is equivalent to 0.11 in terms of $C(60/100)$.

\begin{equation}
\log L_{\rm [CI]}/L_{\rm CO(7-6)} = (0.74 \pm 0.05) - (1.00 \pm 0.06)\,C(60/100).
\end{equation}

The large cross in magenta in the plot is the location of BRI 1335-0417 
from this study, using the measured \CI/CO\,(7$-$6) luminosity ratio and
the $C(60/100)$ value from the SED fit.  The horizontal error bar 
indicates the possible range for its $C(60/100)$ value.  Most of 
the local AGN-dominant LIRGs lie above the solid line. This is likely 
due to the fact that the CO\,(7$-$6) emission is rather insensitive to
AGN gas heating (Lu et al. 2017a).

\begin{figure}
\vspace{-1.0in}
\centering
\includegraphics[width=0.75\textwidth, bb=18 144 592 718]{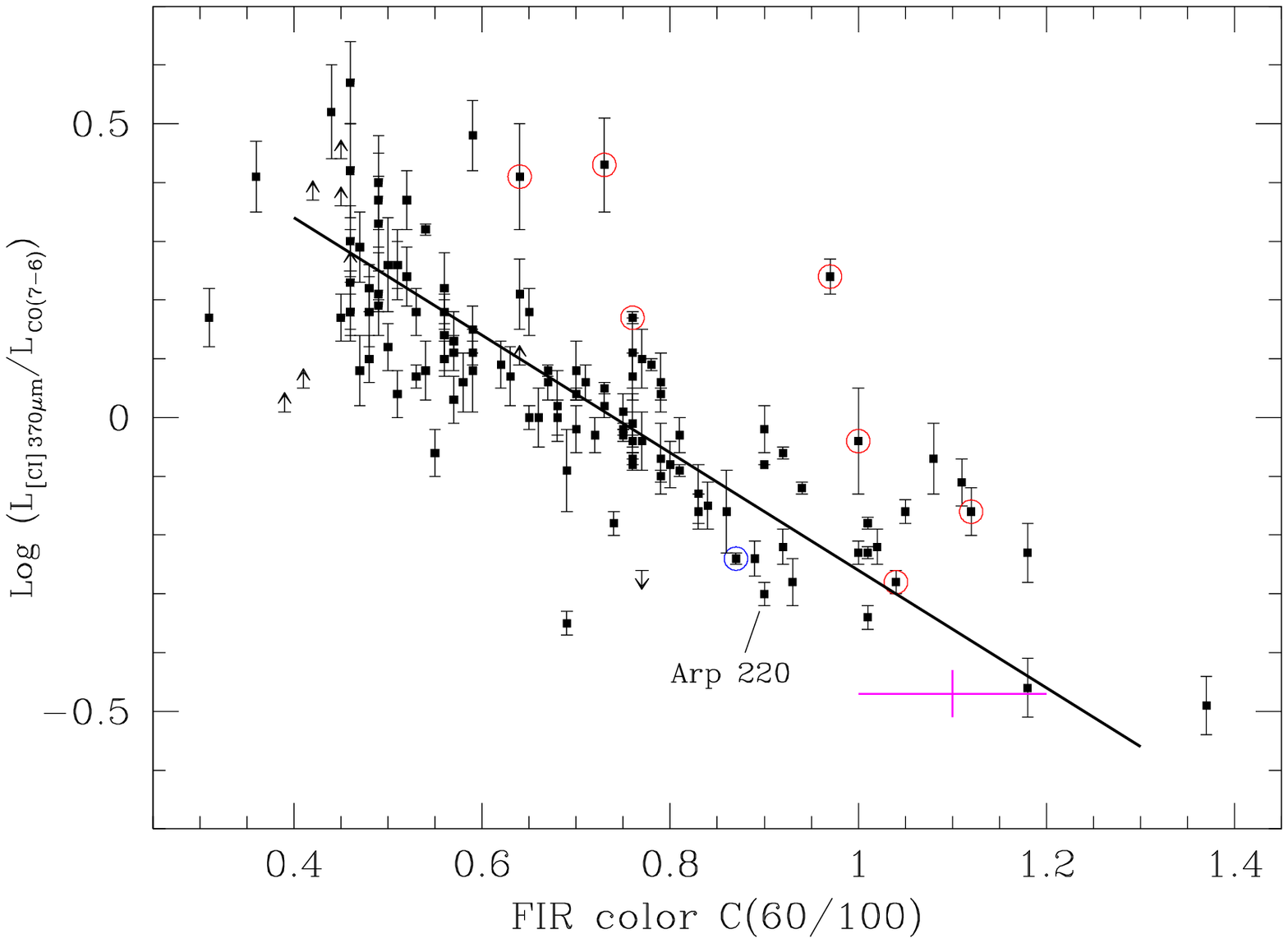}
\vspace{-0.5in}
\caption{Plot of $\log\,L_{\rm [CI]}/L_{\rm CO(7-6)}$ as a function of $C(60/100)$ 
for a flux-limited local sample of LIRGs (filled squares), adopted from Fig. 17 
in Lu et al. (2017a).   The data points circled in red are the galaxies, for which 
the AGN contribution is likely to dominate the bolometric luminosity based on 
a number of mid-IR diagnostics, and the data point circled in blue is NGC 6240 
where there is significant gas heating by shocks unrelated to star formation
(See Lu et al. 2017 for further details).  The solid line is a bisector least-squares
fit to all the data points detected in both \CI\ and CO\,(7$-$6), excluding those 
circled.   The data point in magenta represents BRI 1335-0417 from this study, 
with its horizontal error bar corresponding to the possible range of the $C(60/100)$
value.  Also labelled is Arp 220, one of the brightest local (U)LIRGs.
}
\label{FigA1}
\end{figure}

\end{document}